\documentclass[%
 reprint,
 amsmath,amssymb,
 aps,
]{revtex4-2}

\usepackage{graphicx}
\usepackage{dcolumn}
\usepackage{bm}

\usepackage{slashed}
\usepackage{subcaption}
\usepackage{color}
\usepackage[mathlines]{lineno}

\begin{document}

\preprint{AAPM/123-QED}

\title{\textbf{On the smallness of charm loop effects in $B\to K^{(*)} \ell\ell$ at low $q^2$: light meson Distribution Amplitude analysis}}

\author{Namit Mahajan}
\email{nmahajan@prl.res.in, mahajannamit@gmail.com}
 \affiliation{Theoretical Physics Division, Physical Research Laboratory, Ahmedabad, India - 380009.}
\author{Dayanand Mishra}%
 \email{dayanand.mishra@tifr.res.in}
\affiliation{ 
Department of Theoretical Physics, Tata Institute of Fundamental Research, Mumbai, India - 400001.
}%


\begin{abstract}
The non-local effects originating from the charm quark loops at dilepton invariant masses smaller than the charmonium threshold in  $B\to K \ell\ell$ are evaluated with light meson distribution amplitudes. The revised estimates with B-meson distribution amplitude within a Light Cone Sum Rule approach yielded results about three orders smaller than the original computation. In view of the importance of these non-factorizable soft gluon effects, both conceptually and phenomenologically, an independent evaluation is necessary. It is found that to twist-4 accuracy, these soft gluon effects vanish when evaluated employing the kaon distribution amplitude. Similar results hold for $B\to K^* \ell\ell$ to the leading twist. This eliminates one of the major sources of potential uncertainty which usually makes it difficult for a clear case of new physics, should the data show deviations from the standard model.
\end{abstract}

\maketitle

\date{}

{\bf Introduction:} B-meson decays into leptonic final states, particularly mediated by the neutral current like $B \to K^{(*)}\ell\ell$,  offer a fertile playground to test the Standard Model (SM) of particle physics (for an incomplete list, see \cite{Bobeth:2007dw,Hiller:2014yaa,Kruger:1999xa,Altmannshofer:2008dz,Egede:2008uy,Matias:2012xw,Descotes-Genon:2012isb,Bobeth:2010wg,Lyon:2013gba}). 
Owing to being loop suppressed within the SM, and being relatively cleaner theoretically, these modes have attracted considerable theoretical and experimental attention. 
While the recent experimental results on the lepton flavor universality (LFU) ratios \cite{LHCb:2022vje,LHCb:2022qnv,Smith:2024xgo,CMS:2024syx} agree quite well with the SM predictions \cite{Hiller:2003js,Bordone:2016gaq,Mishra:2020orb,Isidori:2020acz}, it is still not well settled if these modes are being affected by some new physics, albeit not showing up in these LFU ratios. For concreteness, we consider $B \to K\ell\ell$ for the most part, and discuss the $B \to K^*$ decays at the end.

\begin{figure}[h]
	 	\begin{subfigure}{.25\textwidth}
	 		\centering
	 		\includegraphics[width=0.9\linewidth]{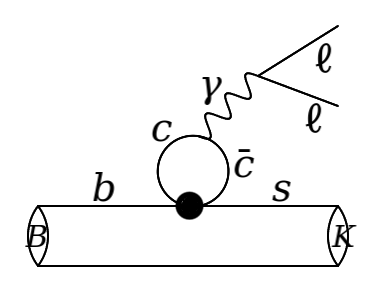}
	 		\caption{}
            \label{leading_charm_loop}
	 	\end{subfigure}%
        \begin{subfigure}{.25\textwidth}
	 		\centering
	 		\includegraphics[width=0.9\linewidth]{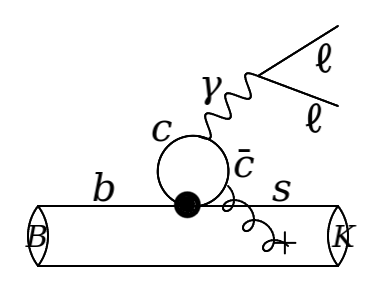}
	 		\caption{}
            \label{soft_g_c_loop}
	 	\end{subfigure}
	 	\caption{(a) Representative Diagram of charm loop contribution to $B \to K\ell\ell$. (b) Typical non-factorizable soft gluon charm loop contribution to $B\to K\ell\ell$.}
	 	\label{fig}
	 \end{figure}

In the SM, the relevant quark level effective Hamiltonian \cite{Buchalla:1995vs} (with $\lambda_i=V_{ib}V_{is}^*$) is:
\begin{eqnarray}
    H_{eff}=&&\frac{4G_F}{\sqrt{2}}\left(\lambda_c \sum_{i=1,2} C_i \mathcal{O}_i + \lambda_t \sum_{i=3,...10} \mathcal{C}_i \mathcal{O}_i\right),
\end{eqnarray}
The relevant operators are 
\begin{eqnarray}
    \mathcal{O}_1 &=&(\Bar{s}_i \gamma_\mu P_L c_j)(\Bar{c}_j\gamma^\mu P_L b_i),\nonumber\\
    \mathcal{O}_2 &=&(\Bar{s}_i \gamma_\mu P_L c_i)(\Bar{c}_j\gamma^\mu P_L b_j)\nonumber,\\
    \mathcal{O}_{7\gamma} &=& -\frac{e m_b}{16\pi^2}(\Bar{s} \sigma_{\mu\nu}m P_R b)F^{\mu\nu},\nonumber\\
    \mathcal{O}_9 &=&\frac{\alpha_{em}}{4\pi}(\Bar{s} \gamma_\mu P_L b)(\Bar{\ell} \gamma^\mu \ell),\nonumber\\
    \mathcal{O}_{10} &=&\frac{\alpha_{em}}{4\pi}(\Bar{s} \gamma_\mu P_L b)(\Bar{\ell} \gamma^\mu \gamma_5 \ell),
\end{eqnarray}
where $i,\ j$ are color indices, $e=\sqrt{4\pi \alpha_{em}}$.  Operators $\mathcal{O}_{7\gamma}$, $\mathcal{O}_9$ and $\mathcal{O}_{10}$ originate from photon and Z penguin diagrams and box diagram. The quark loop contribution (dominantly from the four quark current-current operators, $\mathcal{O}_{1,2}$) is perturbatively computed, including hard gluon QCD contributions, and included along with $C_9$ to define $C_9^{eff}$. Due to the CKM elements involved and a larger mass, the most important contribution stems from the charm quark loop, Fig.(\ref{leading_charm_loop}). The light quark contribution, being CKM suppressed and/or accompanied by small Wilson coefficients, is not discussed here. In the SM, $C_1 = -0.2507$, $C_2 = 1.0136$,    $C_{7\gamma} = -0.3143$, $C_9 = 4.0459$ and $C_{10} = -4.2939$ \cite{Bobeth:1999mk,Bobeth:2003at,Misiak:2004ew,Huber:2005ig,Mahmoudi:2024zna}.
The Wilson coefficients for $\mathcal{O}_{3-6}$ and $\mathcal{O}_{8g}$ are small, and hence these operators are neglected in the discussion below, though these could be analogously included.

	Charm quarks can contribute in one more way: via the charmonium resonances, with the threshold around $q^2 \sim 9$ GeV$^2$: $B\to J/\psi (J/\psi\to \ell\ell) K$ resulting in narrow peaks at the $J/\psi$ and other relevant charmonia masses, with $q^2$ denoting the dilepton invariant mass squared. This contribution is way larger than the short distance contribution due to the effective Hamiltonian, and being non-perturbative in nature, is difficult to model and calculate reliably, but can have important phenomenological consequences (see for example: \cite{Kruger:1996cv,Lyon:2014hpa,Ciuchini:2015qxb,Jager:2014rwa,Chobanova:2017ghn,Bobeth:2017vxj,Hurth:2020rzx,Ciuchini:2020gvn,Ciuchini:2021smi,Isidori:2024lng}). Experimentally, then, a suitable cut is imposed on the dilepton invariant mass to mask out this resonance region. The resonances don't die off very sharply and the tails do extend into the perturbative region. It is thus a common practice to determine the dilepton spectrum in the range, $1< q^2 < 6$ GeV$^2$. The lower limit ensures that the difference in the electron and muon masses is no longer significant, and at the same time avoids the peaking contribution due to the photon pole. This defines the low-$q^2$ region, and it is expected that the theoretical calculations are under control, and thus, reliable in this region.

 The above two types of contributions do not exhaust the charm contribution. There is yet another type of contribution, often dubbed as 'charm-loop effect', which is due to soft gluons from the charm loop having virtuality $|k^2| << 4m^2_c$, Fig.(\ref{soft_g_c_loop}). This case goes beyond the perturbation theory, and could potentially be a leading uncertainty in the theoretical calculations at low-$q^2$. It is, thus, rather important to have a good estimate of such a contribution. To this effect, consider the two current-current operators listed above. These can be combined, after Fierz transformation, into the following combinations: $C_1\mathcal{O}_1+C_2\mathcal{O}_2= \left(C_1+\frac{C_2}{3}\right)\,\mathcal{O} + 2C_2\, \tilde{\mathcal{O}}$, 
where, $ \mathcal{O}=(\bar{c}\Gamma_\mu c)(\bar{s}\Gamma^\mu b),\ \ \text{and}\ \   \tilde{\mathcal{O}}=(\bar{c}\Gamma_\mu T^a c)(\bar{s}\Gamma^\mu T^a b)$.
At the one gluon level, operator $\mathcal{O}$ contributes to the factorizable amplitude while the operator $\tilde{\mathcal{O}}$ leads to a non-factorizable contribution.

We are specifically interested in the region $q^2 << 4m^2_c$. The soft gluon contributions to the charm loop, leading to non-local hadronic matrix elements, have been previously evaluated within the Light Cone Sum Rule (LCSR) approach employing B-meson Light Cone Distribution Amplitudes (LCDAs) \cite{Khodjamirian:2010vf} (referred to as KMPW below). A revised calculation \cite{Gubernari:2020eft} (referred to as GDV), including the sub-leading terms and updated parameter values, found that the non-factorizable soft gluon charm loop contribution is smaller than that evaluated earlier by KMPW, by almost three orders (see \cite{Khodjamirian:2012rm,Gubernari:2022hxn} for an overview of different types of corrections and present status, and \cite{LHCb:2023gel,LHCb:2024onj} for experimental efforts to pinpoint these non-local effects). This is an important conclusion, since, if so, the simple minded use of the short distance Wilson coefficients in calculating observables in the low-$q^2$ region would suffice. Comparison with the experimentally available information seems to suggest an effective shift in the Wilson coefficient $C_9$: $\delta C_9 \sim -1$. The errors are still large and thus, whether this has a hadronic or new physics origin is far from clear, let alone resolved. In this respect, it is important to have an independent evaluation of the above mentioned non-factorizable contributions. To this end, we employ the light meson LCDAs, though, there may be an issue of concern while using light meson DAs. In such a case, possible contamination from light states in the dispersion relations can creep in. This could, however, be handled following the approach suggested in \cite{Khodjamirian:2000mi}. Further, for the present purpose, we set kaon mass to zero: $m_K = 0$, but not when $m_K$ appears as an overall multiplicative factor in the DAs.

The amplitude for the process is written as:
\begin{eqnarray}
        \langle K \ell\ell | H_{eff} |B\rangle &= \frac{\alpha}{4\pi} \frac{4G_F}{\sqrt{2}} V_{cb} V_{cs}^* \Bigg[(C_9 L^\mu_V + C_{10} L^\mu_A)\nonumber\\ 
    &\langle K|\bar{s}\gamma_\mu P_L b|B\rangle-\frac{16\pi^2}{q^2} L^\mu_V \langle K|\mathcal{H}_\mu |B\rangle \Big]
\end{eqnarray}
with,
\begin{eqnarray}
     L^\mu_{V(A)}&=&\bar{\ell}\gamma^\mu (\gamma_5) \ell, \ \text{and}\ \nonumber\\ 
    \mathcal{H}_\mu &=&i\int d^4x e^{iq.x}T\lbrace{ j_\mu^{em}(x),\left(C_1+\frac{C_2}{3}\right)\mathcal{O} + 2C_2 \tilde{\mathcal{O}}\rbrace},\nonumber
\end{eqnarray}
   
leading to the factorized part of the amplitude (which will not be pursued further):
\begin{eqnarray}
    \mathcal{H}_{\mu,\,fac} &=& i\int d^4x e^{iq.x}T\lbrace{ j_\mu^{em}(x),\left(C_1+\frac{C_2}{3}\right)\mathcal{O}\rbrace} \\
    &=& i\left(C_1+\frac{C_2}{3}\right) \frac{9}{16\pi^2} (q_\mu q_\rho-q^2 g_{\mu\rho})h(q^2)\bar{s}\gamma_\rho P_L b\nonumber
    \label{fac-hamil}
\end{eqnarray}
The function $h(q^2)$ is the perturbative contribution which enters $C_9^{eff}$ \cite{Grinstein:1988me,Misiak:1992bc,Buras:1994dj}.

\textbf{Non-factorizable part of the amplitude:}
The non-factorizable charm loop contribution, up to overall factors, is given by the correlator
\begin{eqnarray}
  \mathcal{H}_{\mu,\,non-fac} &\sim& \int d^4x e^{iq.x}T\lbrace{ j_\mu^{em}(x),\tilde{\mathcal{O}}\rbrace} \label{charmloopeq1} \\
  & \propto& \int d^4x e^{iq.x}\langle 0| T\lbrace{ \bar{c}\gamma_\mu c(x),(\bar{c}\gamma_\rho T^a c)(0)\rbrace} |0\rangle \nonumber
\end{eqnarray}

Following \cite{Khodjamirian:2010vf}, one can establish light cone dominance of Eq.(\ref{charmloopeq1}). Consider a small, but finite, time-like $q^2$, and define a unit vector, $v_c=q/\sqrt{q^2}=(1,\vec{0})$, in the rest frame of the virtual photon. The virtual $c$-quark momentum can be decomposed into static and residual four momenta, $p_c=m_c v_c+\tilde{p_c}$, allowing the charm field to be decomposed as $c(x)=e^{-im_cv_c.x} h(x)$, where $h(x)$ contains only the residual momentum components. Now, since the virtual $\bar{c}$-quark four momentum is $p_{\bar{c}}=q-p_c$, the Dirac-conjugated $\bar{c}$-field enters as $\bar{c}(x)=e^{-im_cv_c.x} \bar{h}(x)$. Using these field redefinitions, Eq.(\ref{charmloopeq1}) $\sim \int d^4x e^{-i(2m_c\,v_c -q).x}\langle 0| T\lbrace{ \bar{h}\gamma_\mu h(x),(\bar{h}\gamma_\rho T^a h)(0)\rbrace} |0\rangle.$
This integral gets dominant contributions from the region where the exponent does not oscillate strongly, i.e., $(2m_c v_c-q).x\sim 1$, leading to $
    \langle x^2\rangle\sim \frac{1}{(2m_c v_c-q)^2}\sim\frac{1}{(2m_c-\sqrt{q^2})^2}$.
Hence, for $q^2\ll 4m_c^2$, the dominant region of integration over $x$ in Eq.(\ref{charmloopeq1}) is concentrated near the light-cone $x^2\approx 0$. Furthermore, when $q^2$ grows and approaches the threshold, $4m_c^2$, the Light Cone Operator Product Expansion (LCOPE) becomes invalid. Though a specific frame is chosen here, this analysis is valid in any frame due to Lorentz invariance. Furthermore, it is to be noted that the light cone dominance of the correlation function, Eq.(\ref{charmloopeq1}), should not depend on how the matrix element (correlator) is computed.

Having established the light-cone dominance, we proceed to the evaluation of the soft gluon non-factorizable contributions to the charm loop resulting from the gluon field strength part of the propagator ($x_{12} = x_1-x_2$):
\begin{eqnarray}
    &\langle 0| T\lbrace{ c(x_1)\bar{c}(x_2)\rbrace} |0\rangle =i\int \frac{d^4k}{(2\pi)^4}e^{-ik.x_{12}}\Big[ \frac{\slashed{k}+m_c}{k^2-m_c^2}\nonumber\\ &-\int_0^1 du G^{\mu\nu}(ux_1+\bar{u} x_2) \left(\frac{\bar{u}(\slashed{k}+m_c)\sigma_{\mu \nu}+u\sigma_{\mu\nu}(\slashed{k}+m_c)}{2(k^2-m_c^2)^2}\right)\Big]
\end{eqnarray}
In the rest frame of the decaying $B$ meson, with $v=(1,0,0,0)$, define two light cone vectors $n_{\pm}$ such that
$2v=(n_++n_-), \ n_+^2=n_-^2=0, \ n_+n_-=2, \text{and}\ \ q=(n_-q)\frac{n_+}{2}+(n_+q)\frac{n_-}{2}+q_\perp $
We choose $q_\perp=0$ and consider the kinematical situation when the lepton pair invariant mass is small, i.e., $q^2= (n_+q)(n_-q)<<4m_c^2<<m_b^2$, while $v.q\sim m_b/2$ is large, so that one of the components of $q$ dominates. Specifically, we choose: $(n_+q) \sim {\mathcal{O}}(m_b)$, i.e., $q$ is essentially along $n_+$. It is to be noted that to ensure the convergence of the LCOPE, one requires: $4m_c^2-q^2>>\Lambda m_b$.  

As a next step, we explicitly rederive the charm loop contribution with one soft gluon term included, and use the form for $\tilde{I}_{\mu\rho\alpha\beta}$ that matches with the GDV calculation \cite{Gubernari:2020eft}:
\begin{eqnarray}
    \mathcal{H}_{\mu,\,non-fac} = i\int d^4x e^{iq.x}T\lbrace{ j_\mu^{em}(x),2C_2\tilde{\mathcal{O}}\rbrace}\nonumber\\
    = 2C_2\int d\omega \tilde{I}_{\mu\rho\alpha\beta} \left(\bar{s}\gamma^\rho  P_L\delta\left(\omega- \frac{i n_+ D}{2}\right)G^{\alpha\beta}b\right)
    \label{non-fac_hamil}
\end{eqnarray}
 The function $\tilde{I}_{\mu\rho\alpha\beta}$ is given by
\begin{eqnarray}
    &&\tilde{I}_{\mu\rho\alpha\beta}(q,\omega)=-\int_0^1 du \int_0^1 dt\frac{1}{\Delta}\,\Big[4t(1-t)\Big(\tilde{q}_\mu\epsilon_{\rho\alpha\beta \tilde{q}}\nonumber\\ &&-2u\tilde{q}_\beta\epsilon_{\mu\rho\alpha\tilde{q}}+ 2u \tilde{q}^2\epsilon_{\mu\rho\alpha\beta}\Big) + (1-2 u)\tilde{q}^2 \epsilon_{\mu\rho\alpha\beta}\Big],
\end{eqnarray}
with $\Delta=m_c^2-t(1-t)\tilde{q}^2$. Here $\tilde{q}=q-u\omega n_-$, where the direction $n_-$ corresponds to the gluons emitted antiparallel to $q$, i.e., in the same direction as the $s$-quark (K-meson). In obtaining the above form, $G_{\alpha\beta}(ux)=\exp\big[-iux_{\sigma} (iD^{\sigma})\big]G_{\alpha\beta}(0)$ has been used.

We then sandwich Eq.(\ref{non-fac_hamil}) between the mesonic states, $B$ and $K$, interpolate the $B$ meson, and employ the LCDAs for the $K$ meson along with the free b-quark propagator to obtain (with $\Gamma^{\rho} = \gamma^{\rho}(1-\gamma_5)$)
\begin{eqnarray}
    &&\langle K|{\mathcal{H}}_{\mu,\ non-fac}| B\rangle
    = 2im_bC_2\int d\omega \int \frac{d^4p'}{(2\pi)^4}\int d^4y \nonumber \\ &&e^{i(p'-p_B)\cdot y}\tilde{I}_{\mu\rho\alpha\beta} \langle K|T\lbrace\bar{s}\Gamma_\rho \delta\left(\omega- \frac{i n_+ D}{2}\right)G^{\alpha\beta}(0)\nonumber\\ &&\frac{\slashed{p'}+m_b}{p'^2-m_b^2}\gamma_5 d (y)\rbrace| 0\rangle \ \
    \end{eqnarray}
Making use of the equation: $ \Gamma^\rho (\slashed{p'}+m_b)\gamma_5=p'^\rho(1+\gamma_5)-ip'_\eta\sigma^{\rho\eta}(1+\gamma_5)-m_b \gamma^\rho (1-\gamma_5)$, the non-factorizable contribution to twist-3 accuracy, when employing the K-meson DA (see for example \cite{Ball:2004ye,Belyaev:1993wp}), reads
\begin{eqnarray}
    &&\langle K|{\mathcal{H}}_{\mu,\ non-fac}| B\rangle|_{twist-3} = 2m_bC_2\int d\omega \int \frac{d^4p'}{(2\pi)^4} \nonumber\\ &&\int d^4y \frac{e^{i(p'-p_B)\cdot y}}{p'^2-m_b^2}\tilde{I}_{\mu\rho\alpha\beta}p'_\eta \langle K|T\lbrace\bar{s} \delta\left(\omega- \frac{i n_+ D}{2}\right)\nonumber\\ &&\sigma_{\rho\eta}\gamma_5G^{\alpha\beta}(0) d (y)\rbrace | 0\rangle \nonumber\\
    &&=2im_bC_2 f_{3k}\int d\omega\int d\alpha_s d\alpha_d  \frac{1}{(p_B-\alpha_d p_K)^2-m_b^2} \int du  \nonumber\\ &&\int dt \frac{8t(t-1)u}{\Delta} \left(\epsilon^{\mu\rho\alpha\beta} p_{K\rho}\tilde{q}_\alpha p_{B\beta}\right) (\tilde{q}.p_K)\, \phi_{3k}(\alpha_i,\mu) \nonumber\\ &&\delta(\alpha_s+\alpha_d+ \omega-1) \label{nonfactw3}
\end{eqnarray}
Now, consider the term $\epsilon^{\mu\rho\alpha\beta} p_{K\rho}\,\tilde{q}_\alpha\, p_{B\beta}$, and use the four momentum conservation, $p_{B\mu}=p_{K\mu}+q_{\mu}$, to reduce it to $\epsilon^{\mu\rho\alpha\beta} p_{K\rho}\, \tilde{q}_\alpha\, q_{\beta}$. Recall that $\tilde{q}=q-u\omega n_-$, where $n_-$ is in the same direction as the $s$ quark (or the $K$ meson) in the $B$ meson rest frame. The antisymmetry of Levi-Civita then leads to the vanishing result of $\langle K|{\mathcal{H}}_{\mu,\ non-fac}| B\rangle$ at twist-3.

Alternatively, one could first compute the matrix element for three particle contributions, and then proceed to the loop computation. In such a case, one would then start with the following expression (overall factors are suppressed for convenience)
\begin{eqnarray}
  &&{\mathrm{Nonfac Amp}} \sim \int d^4x\, e^{iq.x}\int \frac{d^4k}{(2\pi)^4} \int \frac{d^4k'}{(2\pi)^4} e^{i(k'-k).x} \nonumber\\ &&\int du \Big[\frac{\bar{u}\slashed{k'}\sigma_{\alpha\beta}+u\sigma_{\alpha\beta}\slashed{k'}+m_c\sigma_{\alpha\beta}}{2(k'^2-m_c^2)^2}\Big]\gamma_\mu\frac{\slashed{k}+m_c}{k^2-m_c^2}\Gamma_\rho \nonumber\\ &&\otimes \int d^4y\, e^{ip_K.y}\int \frac{d^4p'}{(2\pi)^4} e^{-ip'.y}\bigg(\Big[\Gamma_\rho\frac{\slashed{p'}+m_b}{p'^2-m_b^2}\gamma_5\Big]_{ab} \nonumber\\ && \langle K|T\lbrace \bar{s}_a(0)G^{\alpha\beta}(ux)d_b(y)\rbrace |0\rangle \bigg)
\end{eqnarray}
After some Dirac algebra, and focusing on the relevant $\sigma_{\rho\eta}$ term, the matrix element reads
\begin{eqnarray}
   && -ip'_\eta \langle K|T\lbrace \bar{s}(0)\, \sigma_{\rho \eta}\, \gamma_5\, G^{\alpha\beta}(ux)\, d(y)\rbrace |0\rangle =p'_\eta f_{3k} \nonumber \\ && \Big[ (p_{K \alpha}\, p_{K \mu}\, g_{\beta\nu}-p_{K\beta}\, p_{K\mu}\, g_{\alpha\nu})-(p_{K\alpha}\, p_{K\nu}\, g_{\beta\mu}-p_{K\beta}\, p_{K\nu}\, g_{\alpha\mu})\Big]\nonumber\\ &&\otimes \int d\alpha_s\, d\alpha_d\, d\alpha_g\, \phi_{3k} (\alpha_i,\mu) e^{i\alpha_d\, p_K\cdot y}e^{iu\, \alpha_g\, p_K \cdot x } \nonumber\\ && \delta(1-\alpha_s-\alpha_d-\alpha_g)
\end{eqnarray}
Looking at the structure, and the fact that two ways of computing the given matrix element should ideally yield the same result, one would guess that $\alpha_g$ here is the same or analogous to $\omega$ in Eq.(\ref{nonfactw3}). It remains to be seen if this is indeed the case.

Using the large component of $p_K$, one can reduce the exponent: $
    u\alpha_g\, (p_K\cdot x) \rightarrow u\alpha_g \frac{m_b}{2}n_-\, x$. This, then followed by a change of variable from $\alpha_g m_b/2$ to $\omega$, has the effect:
$\int d\alpha_g e^{iu\, \alpha_g\, (p_K\cdot x)}\to \frac{2}{m_b} \int d\omega e^{iu\omega\, n_- \,x}$.

Plugging back, the original expression that one started with now has the form as the one obtained when the charm loop contribution was calculated by translating the gluonic strength tensor $G_{\alpha\beta}(ux)$.
This then confirms the expectation that $\omega$ in the manipulations above is essentially $\alpha_g$ up to a constant factor. Or thinking differently, the delta-function over $\omega$, generated via the use of the translation operator, will eventually end up restoring the argument of the gluon field strength tensor.

Proceeding either way, one arrives at the result that to twist-3 accuracy in the K-meson DA, the non-factorizable soft gluon charm loop contribution vanishes. This would have been the dominant contribution, save for some specific reason, and is, thus, already indicative of the smallness of the non-factorizable charm loop contribution found in the GDV calculation.

Similarly, the twist-4 contribution is found to be:
\begin{eqnarray}
    &&\langle K|{\mathcal{H}}_{\mu,\ non-fac}| B\rangle|_{twist-4} = -m_b C_2 f_{k} \int d\omega \ \ \nonumber\\ && \int d\alpha_s d\alpha_d \int \frac{d^4p'}{(2\pi)^4}\int d^4y  \frac{e^{i(p'-p_B+\alpha_d p_K)\cdot y}}{p'^2-m_b^2}\int du \nonumber\\ && \int dt \frac{8t(t-1)u}{\Delta} \left(\frac{(p_K.\tilde{q})}{p_K.y}\epsilon^{\mu\rho\alpha\beta} p_{K\rho}\tilde{q}_\alpha y_\beta\right)  \nonumber\\ &&\delta(\alpha_s+\alpha_d+ \omega-1) (\phi_{\parallel}(\alpha_i,\mu)+\phi_{\perp}(\alpha_i,\mu))
\end{eqnarray}
With $p'-p_B+\alpha_d p_K = p''$, and $f(p_K)= \epsilon^{\mu\rho\alpha\beta}\, p_{K\rho}\,\tilde{q}_\alpha \,p_{K\eta}\,\tilde{q}_\xi \,g^{\xi\eta}$, consider the term in the parenthesis, along with integration over $y$: $\int d^4y e^{ip''\cdot y} \frac{y_\beta}{p_K.y} f(p_K)$.
After some straightforward manipulations (under the integral sign), this term can be cast in a form which has momentum derivative acting on $f(p_K)$, which is trivially zero: $\frac{\partial}{\partial p_K^\beta}f(p_K)=\epsilon^{\mu\rho\alpha\beta}g^{\xi\eta} (g_{\rho \beta} p_{K\eta}+g_{\eta\beta} p_{K\rho})\tilde{q}_\alpha\tilde{q}_\beta = 0$.
    It then leads to the vanishing result of $\langle K|{\mathcal{H}}_{\mu,\ non-fac}| B\rangle$ even at twist-4.

Hence, one can conclude that when employing light meson DAs, to twist-4 accuracy,
\begin{eqnarray}
    \langle K|{\mathcal{H}}_{\mu,\ non-fac}| B\rangle|_{twist-3 + twist-4}  = 0 
\end{eqnarray}
This is the main result of this work. While in the case when B-meson LCDAs are employed, the nonfactorizable charm loop contribution was found to be small, in the present case when we employ kaon LCDAs, the contribution turns out to vanish both at twist-3 and twist-4 levels. Higher twist contributions are expected to be small. Both for twist-3 and twist-4 contributions, the vanishing result was a consequence of Levi-Civita eventually being contracted to a symmetric quantity.  This argument, thus, provides us with a rough basis for what is expected from such a computation. And this indeed seems to be borne out when the calculation of such effects is carried out in a different way, i.e., employing B-meson LCDAs. However, in that case, the end result of having very small charm loop effect emerges only after consistently including the sub-leading terms and performing a complete calculation. In contrast, in the case when light meson DAs are employed and B-meson is interpolated, the contractions simply yield a zero result, without actually having to set up a dispersion relation. This, then, also avoids possible complications that could arise due to the use of light meson DAs, as alluded to above.  

We also consider similar charm induced non-local effects in $B \to K^*$ mode to twist-3 accuracy. Focusing on the $\sigma_{\rho\eta}$ term, the relevant part coming from the matrix element \cite{Ball:2004rg} contracting with  $\tilde{I}_{\mu\rho\alpha\beta}$ has the form
\begin{eqnarray}
   &&  \langle K^*(p_K,\lambda)|T\lbrace \bar{s}(0)\, \sigma_{\rho \eta}\, \, G^{\alpha\beta}(ux)\, d(y)\rbrace |0\rangle \propto \frac{e^{(\lambda)}y}{2p.y}\nonumber\\ &&   \Big[ (p_{K \rho}\, p_{K \alpha}\, \tilde{g}_{\sigma\beta}- p_{K\rho}\, p_{K\beta}\, \tilde{g}_{\sigma\alpha}\big) -\big(\rho \longleftrightarrow \sigma\big)\Big], \nonumber
\end{eqnarray} 
where $\tilde{g}_{\alpha\beta}= g_{\alpha\beta} - (p_{K \mu} y_\nu + p_{K \nu} y_\mu)/(p_K.y)$. It is then straightforwardly found that
\begin{eqnarray}
    \langle K^*|{\mathcal{H}}_{\mu,\ non-fac}| B\rangle|_{twist-3}  = 0. 
\end{eqnarray}
This shows that the dominant contribution is zero and therefore even for $B \to K^*$ decays, non-factorizable charm loop effects are going to be small.

{\bf Results and Discussion:}
Theoretical calculations, even for semi-leptonic modes like $B\to K^{(*)} \ell\ell$ at low $q^2$ where they are expected to be clean and reliable, suffer from uncertainties beyond those stemming from hadronic form-factors or other input parameters. More specifically, for such modes, the non-factorizable soft gluon contributions via the charm loops need to be understood well. The revised estimates \cite{Gubernari:2020eft} found that these effects are almost three orders smaller than the originally computed \cite{Khodjamirian:2010vf}. Both of these studies had employed B-meson LCDAs in their computations. An independent check, say using light meson LCDAs, computing the same quantity would bring a better theoretical understanding. In the present work, we have computed these non-factorizable contributions and found them to vanish up to twist-4 for $B\to K \ell\ell$. Corrections due to non-zero kaon mass and higher twists are expected to be small. Since, in the present way of computing, the vanishing result is obtained without the need to set-up a dispersion relation and making use of quark-hadron duality, this result shows that the smallness of these charm loop effects is to be expected. A rough way to argue in favour of this is to realise that the loop computation yields a function built out of Levi-Civita tensors, which then contracts with a highly symmetric function coming from the distribution amplitudes of the light meson. This has been explicitly verified in two ways and a rough correspondence is also established between the integration variables in two ways of calculating. Similar conclusions are also reached for $B\to K^* \ell\ell$ modes.
We can, then, practically conclude that the non-factorizable charm-loop effect can be safely neglected. This will have important phenomenological implications. As more data is accumulated, and if the data in specific bins (where these charm loop effects could have made a significant contribution) exhibit tension with the SM expectations, the case for an underlying new physics origin will strengthen many folds. Also, in the new physics analyses and fits, such contributions need not be included, and a direct use of short distance Wilson coefficients can be made.

\begin{acknowledgments}
This work is supported by the Department of Space (DOS), Government of India. NM acknowledges partial support under the MATRICS project (MTR/2023/000442) from the Science \& Engineering Research Board (SERB), Department of Science and Technology (DST), Government of India. DM acknowledges the support from Physical Research Laboratory, where the work was initiated.
\end{acknowledgments}

\nocite{*}
\bibliography{BKcharm}

\end{document}